\documentclass[a4paper,fleqn,usenatbib]{mnras}
\usepackage[T1]{fontenc}
\usepackage{ae,aecompl}

\usepackage{graphicx}	
\usepackage{amssymb}	

\usepackage{multicol}
\usepackage[usenames]{color}
\usepackage{amsmath}
\usepackage{enumitem}
\usepackage{bm}

\newcommand {\bc}{\begin {center}}
\newcommand {\ec}{\end {center}}
\newcommand {\be}{\begin {equation}}
\newcommand {\ee}{\end {equation}}
\newcommand {\beq}{\begin {eqnarray}}
\newcommand {\eeq}{\end {eqnarray}}

\newcommand {\ergs}{{\rm erg\ \rm s^{-1}}}
\newcommand {\comment}[1]{}

\def\lbar {\lambda\hskip-5pt\raise3pt\hbox {--}}
\def\lbr {\lambda\raise2pt\hbox {\hskip-4pt{\scriptsize --}}_\C}

\renewcommand{\d}{{\rm d}}

\renewcommand{\d}{{\rm d}}

\newcommand{\vect}[1]{\boldsymbol{\mathbf{#1}}}
\newcommand{\red}[1]{\textcolor{black}{#1}}

\title[Unstable magnetospheric accretion flows]
{Magnetospheric Flows in X-ray Pulsars I :\\ 
Instability at super-Eddington regime of accretion}
\author[A. A.~Mushtukov et al.] 
{A.~A.~Mushtukov,$^{1}$\thanks{E-mail: alexander.mushtukov@physics.ox.ac.uk (AAM)}  
A. Ingram,$^{2}$
V.~F.~Suleimanov,$^{3}$
N. DiLullo,$^{4}$ 
M. Middleton,$^{5}$
\newauthor
S.~S.~Tsygankov,$^{6}$
M. van der Klis,$^{7}$
S. Portegies Zwart$^{8}$
  \\ 
$^1$ Astrophysics, Department of Physics, University of Oxford, Denys Wilkinson Building, Keble Road, Oxford OX1 3RH, UK\\
$^2$ School of Mathematics, Statistics and Physics, Newcastle University, Herschel Building, Newcastle Upon Tyne, UK \\
$^3$Institut f\"{u}r Astronomie und Astrophysik, Kepler Center for Astro and Particle Physics, Universit\"{a}t T\"{u}bingen, \\ Sand 1, 72076 T\"{u}bingen, Germany\\
$^4$ Thacher School, 5025 Thacher Road, Ojai, CA 93023-8304, USA \\
$^5$ Department of Physics and Astronomy, University of Southampton, Highfield, Southampton SO17 1BJ, UK \\
$^6$ Department of Physics and Astronomy,  FI-20014 University of Turku, Finland \\
$^7$ Anton Pannekoek Institute, University of Amsterdam, Science Park 904, 1098 XH Amsterdam, The Netherlands \\
$^8$ Leiden Observatory, Leiden University, NL-2300RA Leiden, The Netherlands \\
} 

\pubyear{2023}

\begin{document}
\label{firstpage}
\pagerange{\pageref{firstpage}--\pageref{lastpage}}
\maketitle

\begin{abstract}
Within the magnetospheric radius, the geometry of accretion flow in X-ray pulsars is shaped by a strong magnetic field of a neutron star. 
Starting at the magnetospheric radius, accretion flow follows field lines and reaches the stellar surface in small regions located close to the magnetic poles of a star.
At low mass accretion rates, the dynamic of the flow is determined by gravitational attraction and rotation of the magnetosphere due to the centrifugal force. 
At the luminosity range close to the Eddington limit and above it, the flow is additionally affected by the radiative force. 
We construct a model simulating accretion flow dynamics over the magnetosphere, assuming that the flow strictly follows field lines and is affected by gravity, radiative and centrifugal forces only.
The magnetic field of a NS is taken to be dominated by the dipole component of arbitrary inclination with respect to the accretion disc plane.
We show that accretion flow becomes unstable at high mass accretion rates and tends to fluctuate quasi-periodically with a typical period comparable to the free-fall time from the inner disc radius.
The inclination of a magnetic dipole with respect to the disc plane and strong anisotropy of X-ray radiation stabilise the mass accretion rate at the poles of a star, but the surface density of material covering the magnetosphere fluctuates even in this case.
\end{abstract}

\begin{keywords}
accretion, accretion discs -- magnetic fields -- stars: neutron  -- stars: oscillations -- X-rays: binaries
\end{keywords}

\section{Introduction}
\label{sec:Intro}

X-ray pulsars (XRPs) are accreting strongly magnetised neutron stars (NSs) in compact binary systems (see  \citealt{2022arXiv220414185M} for recent review).
Magnetic field at the NS surface in XRPs confirmed by detection of the cyclotron lines \citep{2019A&A...622A..61S} is known to be of the order of $10^{12}\,{\rm G}$ or even stronger.
Large scale magnetic field is expected to be dominated by the dipole component (that can be distorted by the accretion process, see \citealt{2014EPJWC..6401001L} for review), but in some specific cases there was reported possibility of the field heavily contributed by non-dipole components (see, e.g., \citealt{2017A&A...605A..39T,2017Sci...355..817I,2022MNRAS.515..571M}).
Observed luminosity of XRPs covers more then seven orders of magnitude from $10^{32}\,\ergs$ up to $10^{41}\,\ergs$.
The lower edge of accretion luminosity range is partly related to the onset of the ``propeller" state of accretion when the centrifugal barrier due to the rotating NS magnetosphere prevents accretion onto the stellar surface \citep{1975A&A....39..185I,1998A&A...340...85R,2006ApJ...646..304U}.
The brightest XRPs belong to the class of recently discovered pulsating ultra-luminous X-ray sources (ULXs, \citealt{2014Natur.514..202B,2017Sci...355..817I,2016ApJ...831L..14F,2017MNRAS.466L..48I,2018MNRAS.476L..45C} and \citealt{2021AstBu..76....6F,2023NewAR..9601672K} for review).
The apparent luminosity of the brightest ULX pulsar - NGC5907~X-1 - exceeds $10^{41}\,\ergs$ \citep{2017Sci...355..817I}.
The relation between the actual and apparent luminosity of bright XRPs is still debated by the community. 
\red{
Specific geometry of X-ray emission region in close proximity to a NS surface does not affect significantly the apparent luminosity of bright XRPs (see, e.g., \citealt{2024MNRAS.527.5374M}).
}
However, there are arguments in favor of a large difference between actual and apparent luminosity due to the geometrical beaming at high mass accretion rates \citep{2009MNRAS.393L..41K,2017MNRAS.468L..59K,2020MNRAS.494.3611K} expected because of possibly strong radiation driven outflows launched from accretion discs \citep{1973A&A....24..337S,2007MNRAS.377.1187P,2009MNRAS.393L..41K,2014ApJ...796..106J}. 
At the same time it has been shown that strong beaming of X-ray radiation is inconsistent with large pulsed fraction observed in six ULX pulsars known up to date, which says in favour of relatively small difference between the actual and apparent luminosity \citep{2021MNRAS.501.2424M,2023MNRAS.518.5457M}.

\begin{figure*}
\centering 
\includegraphics[width=12.cm]{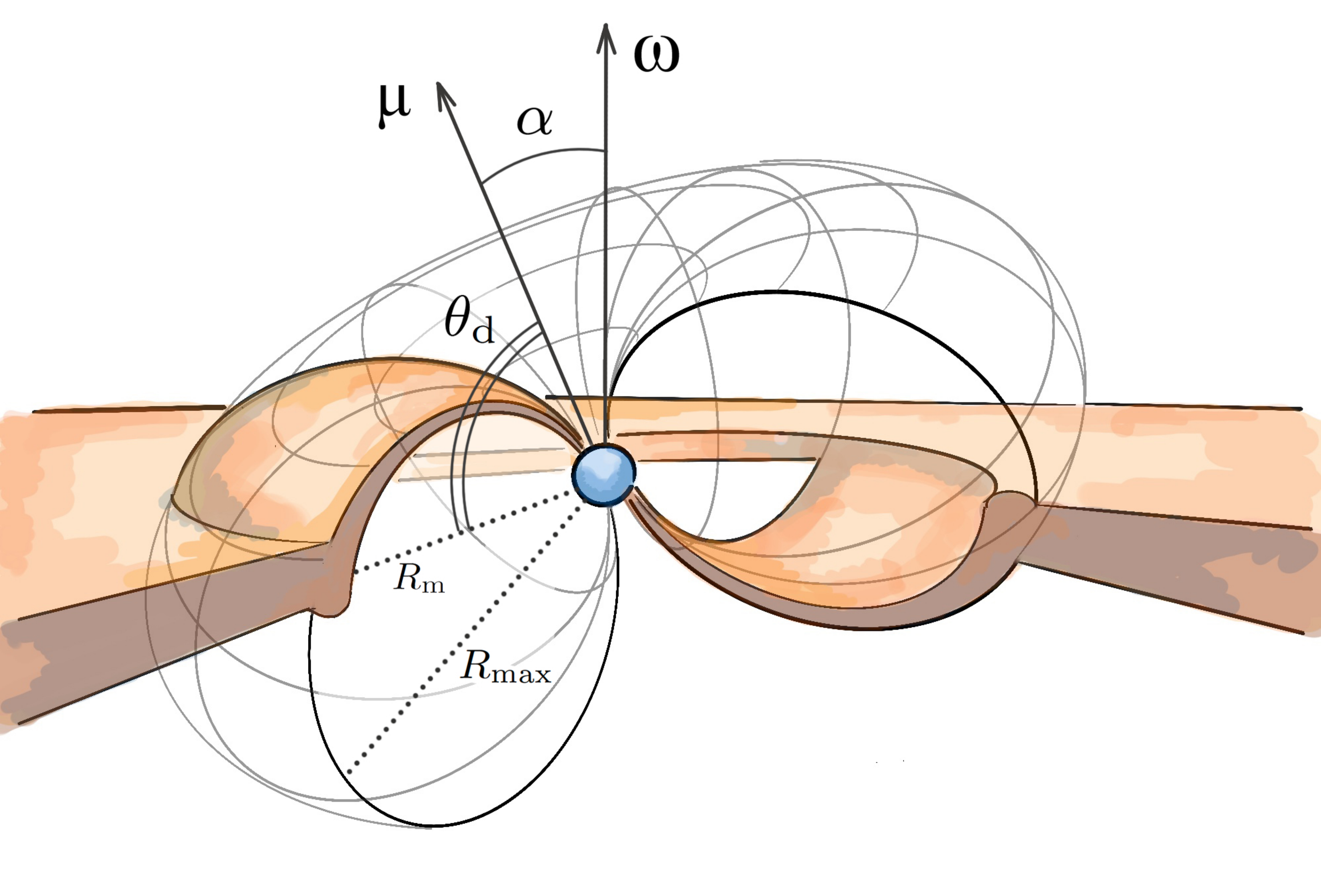}
\caption{
Schematic illustration of accretion geometry onto strongly magnetised NS.
Magnetic field is considered to be dominated by the dipole component, while the rotation axis is assumed to be orthogonal to accretion disc plane.
}
\label{pic:scheme}
\end{figure*}

Accretion flow in the systems hosting strongly magnetised NSs is truncated by magnetic field of a NS at the magnetospheric radius $R_{\rm m}$, which is determined by the NS magnetic field strength, mass accretion rate at $R_{\rm m}$ and geometry of the accretion flow (i.e., disc/wind accretion, disc thickness and physical conditions in it, see \citealt{2019A&A...626A..18C} for detailed discussion).
At the magnetospheric radius, the flow penetrates into the magnetosphere due to the instabilities, including magnetic Rayleigh-Taylor \citep{1976ApJ...207..914A,2008MNRAS.386..673K} and Kelvin-Helmholtz  instabilities \citep{1983ApJ...266..175B}.
\red{Development of the instabilities results in formation of a boundary layer between the disc and magnetosphere of a NS \citep{2014EPJWC..6401001L}.
In the boundary layer, the angular velocity of plasma experiences a transition from the Keplerian rate to the NS spin rate.
}

Within the magnetospheric radius, the flow follows magnetic field lines and finally reaches the surface of a NS within small regions (area $\sim 10^{10}\,{\rm cm^2}$, see Section 4.3 in \citealt{2022arXiv220414185M}) located close the magnetic poles of a star.
Depending on the mass accretion rate and corresponding energy release, the flow is stopped due to the Coulomb collisions in the atmosphere of a NS (expected for $\dot{M}\lesssim 10^{17}\,{\rm g\,s^{-1}}$, see, e.g., \citealt{1969SvA....13..175Z,1995ApJ...438L..99N}) or at the radiation pressure dominated shock above NS surface (expected for the case of $\dot{M}\gtrsim 10^{17}\,{\rm g\,s^{-1}}$, see \citealt{1976MNRAS.175..395B,1981A&A....93..255W,2022MNRAS.515.4371Z}). 

The dynamics of accretion flow in between the magnetospheric radius and NS surface is determined by the geometry of magnetic field lines, gravitational force, centrifugal force and radiative force, which comes into play at sufficiently high mass accretion rates. 
Stable accretion flow covering the magnetosphere of a NS is expected to be optically thick at high mass accretion rates $\dot{M}\gtrsim {\rm few}\times 10^{18}\,{\rm g\,s^{-1}}$ typical for ULXs pulsars and bright Be~XRPs \citep{2011Ap&SS.332....1R} at the peaks of their outbursts (see, e.g., \citealt{2017MNRAS.467.1202M,2019MNRAS.484..687M}).
The optically thick envelope might strongly affect the key observational properties of bright XRPs, including their energy spectra, X-ray polarisation, pulse profiles and timing properties of aperiodic variability.

This paper presents the first simulations of accretion flow dynamics between the inner disc radius and NS surface. 
We account for the geometry of accretion flow shaped by a magnetic field of a NS, and the influence of gravitational, centrifugal and radiative forces. 
The stellar magnetic field is assumed to be purely dipole with a given arbitrary inclination of the magnetic dipole with respect to the accretion disc plane.

\section{Model set up}
\label{sec:Model}

\begin{figure}
\centering 
\includegraphics[width=8.4cm]{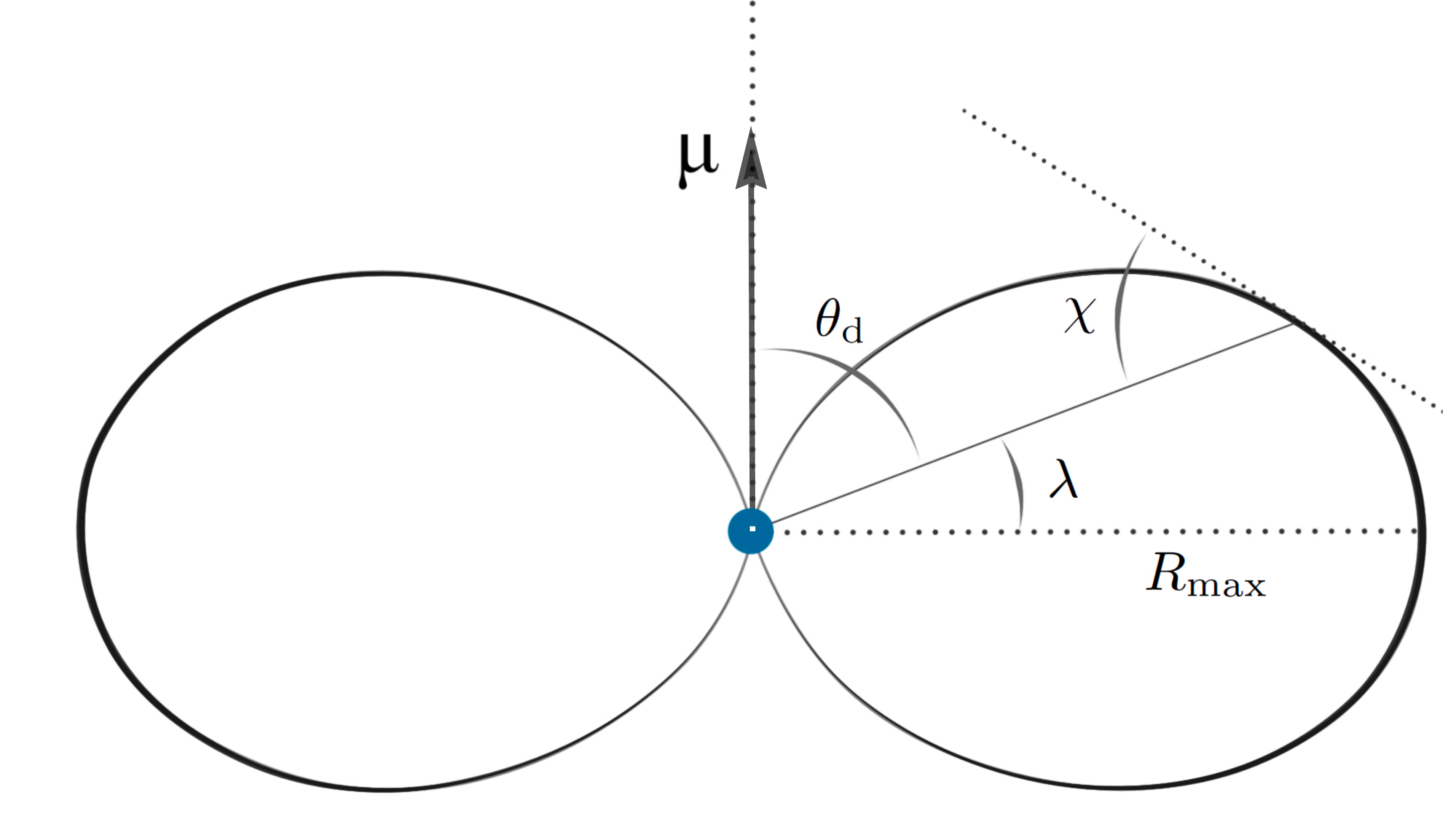}
\caption{
The schematic geometry of dipole magnetic field line, where $R_{\rm max}$ is the linear scale describing magnetic field line, $\lambda$ ($\theta_{\rm d}$) is the latitude (co-latitude) of a point, and $\chi$ is the angle between the position point and the tangent line to the field line (\ref{eq:chi}).  
}
\label{pic:scheme_2}
\end{figure}

\subsection{Geometry of accretion flow}

In this paper, we consider a particular case of accretion from the disc, i.e. the material starts its motion towards the poles of a NS from a specific regions, where the disc is interrupted by the $B$-field.
We assume that the rotational axis of a NS is orthogonal to accretion disc plane, which is likely a case in strongly magnetised accreting NSs.
The disc in XRPs is interrupted at the magnetospheric radius $R_{\rm m}$, which is dependent on the mass accretion rate, NS magnetic field strength and structure.  
In the case of the field dominated by the dipole component, $R_{\rm m}$ can be estimated as
\beq\label{eq:Rm}
R_{\rm m} \approx 1.8\times 10^8\,\Lambda B_{12}^{4/7}\dot{M}_{17}^{-2/7}m^{-1/7}R_6^{12/7}\,\,{\rm cm},
\eeq
where $\Lambda<1$ is a factor depending on the accretion flow geometry (see, e.g., Chapter 6.3 in \citealt{2002apa..book.....F}, and \citealt{2019A&A...626A..18C} for recent discussion related to the case of high mass accretion rates), $B_{12}$ is a surface magnetic field strength in units of $10^{12}\,{\rm G}$, 
$\dot{M}_{17}$ is mass accretion rate in units of $10^{17}\,{\rm g\,s^{-1}}$, 
$m$ is a mass of a NS $M$ in units of $M_\odot$, and $R_6$ is the NS radius in units of $10^6\,{\rm cm}$.
\red{
The effective temperature in accretion disc can be estimated as
\beq\label{eq:Teff_1}
\sigma _{\rm SB}T^4_{\rm eff}=
\frac{3}{8\pi}\frac{GM\dot{M}}{r^3}
\left[1-\beta\left(\frac{R_{\rm m}}{r}\right)^{1/2}\right],
\eeq 
where $\sigma_{\rm SB}$ is the Stefan-Boltzmann constant and $r$ is a distance from a central compact object, and $\beta\in[0;1]$ is a coefficient determined by a distance where stress disappears ($\beta=1$ in the case of stress disappearance at $R_{\rm m}$, while $\beta=0$ corresponds to stress disappearance at $r\ll R_{\rm m}$).
Combining (\ref{eq:Rm}) and (\ref{eq:Teff_1}) we can estimate roughly the effective temperature at the inner disc radius as
\beq\label{eq:Teff_2}
T_{\rm eff}\lesssim 
0.03\,B_{12}^{-3/7}\dot{M}_{17}^{13/28}\,\,\,
{\rm keV}.
\eeq 
The temperature of plasma settling magnetic field lines at the inner disc radius is probably higher because (\ref{eq:Teff_1}) and (\ref{eq:Teff_2}) do not account for the interaction between the magnetosphere of a NS and accretion disc.
}

We assume that the accretion flow within the magnetospheric radius cannot move across magnetic field lines.
In the spherical coordinates $(r,\theta,\varphi)$, the geometry of a dipole field line is given by 
\beq\label{eq:dip_fl}
r=R_{\rm max}\cos^2 \lambda=R_{\rm max}\sin^2 \theta, 
\eeq 
where $R_{\rm max}$ is a linear scale describing dipole magnetic field line,
$\lambda$ is the latitude, and $\theta=\pi/2-\lambda$ is corresponding co-latitude in the reference frame related to the $B$-field.
The linear scale of a field line $R_{\rm max}$ in (\ref{eq:dip_fl}) is typically assumed to be equal to $R_{\rm m}$, but it is a case of magnetic dipole aligned with the disc only.
In the case of inclined magnetic dipole
\beq 
R_{\rm max} = R_{\rm m}\sin^{-2}\theta_{\rm d},
\eeq 
where 
\beq 
\theta_{\rm d}={\rm atan}(\sin^{-1}\varphi \tan^{-1}\alpha)
\eeq  
is the co-latitude, where the disc plane crosses the dipole surface, $\alpha$ is the magnetic obliquity, i.e. the angle between the rotational axis of the NS and the magnetic dipole axis (see Fig.\,\ref{pic:scheme}).
Assuming the inner disc radius independent of the azimuthal coordinate, we get $R_{\rm max}$ dependent on $\varphi$ and magnetic obliquity $\alpha$.

The area on the dipole surface $S$ is related to the latitude $\lambda$ and the azimuthal angle $\varphi$ as
\beq
\d S=R_{\rm max}^2 \cos^4\lambda (1+3\sin^2\lambda)^{1/2} \d\lambda\d\varphi,
\eeq
while the length along the field lines $x$ is related to the latitude as
\beq
\d x = R_{\rm max}\cos\lambda (1+3\sin^2\lambda)^{1/2} \d\lambda.
\eeq
The angle between the position vector of the point at the dipole surface and the tangent line to the dipole magnetic field line (see Fig.\,\ref{pic:scheme_2}) is
\beq\label{eq:chi}
\chi={\rm atan}[0.5\,{\rm tan}^{-1}\lambda].
\eeq

{
Further in this paper we present maps describing the distribution of different quantities (accelerations, surface density, etc.) over the dipole surface of a NS magnetosphere. 
The position of a point on the dipole surface ($R_{\rm max}$ is fixed) is uniquely given by two angles: $\lambda$ (or $\theta$) and $\varphi$. 
To represent magnetospheric maps, we apply Aitoff transformation to the coordinates $\lambda$ and $\varphi$, as is standardly done to represent maps of spherical surfaces:
\beq 
x_{\rm pr}=\frac{2\varrho\sin(\lambda/2)\cos\varphi}{\sin\varrho},
\quad\quad 
y_{\rm pr}=\frac{\varrho\sin\varphi}{\sin\varrho},
\eeq 
where 
$$\varrho = {\rm arccos}\left(\cos\frac{\lambda}{2}\cos\varphi\right).$$
}


\subsection{Dynamics of accretion flow}
\label{sec:Dynamics}

We consider the accretion flow moving under the influence of three forces: the gravitational, the radiative and the centrifugal force.
We assume that the flow covering the magnetosphere of a NS is supersonic and do not account for the influence of internal gas pressure gradients.

The gravitational force is directed towards the compact object,
and the corresponding acceleration along the dipole field line is given by 
\beq
\left|a_{{\rm grav},||}\right|=\cos\chi\,\frac{GM}{r^2}\simeq 1.328\times 10^{10}\,m r_8^{-2}\cos\chi \,\,{\rm cm\,s^{-2}}.
\eeq
The gravitational acceleration along the field lines turns to zero at the equator. 
At $\lambda>0$ ($\lambda<0$), the acceleration $a_{{\rm grav},||}$ is positive (negative). 
Its absolute value of the acceleration increases towards to the central object. 

The centrifugal force is directed perpendicularly to the rotational axis of a NS and dependent on the NS spin period and specific point on the magnetospheric surface.
In the particular case of magnetic dipole aligned with the rotational axis, the corresponding centrifugal acceleration along the field lines is given by
\beq\label{eq:acc_centr_1}
a_{{\rm cen},||}(\alpha=0)=\omega^2 R_{\rm m}\cos^3\lambda \cos(\chi-\lambda),
\eeq
where $\omega$ is the angular velocity of a NS. 
In a general case of inclined magnetic dipole, (\ref{eq:acc_centr_1}) has to be calculated as
\beq\label{eq:a_cent_gen}
a_{{\rm cen},||} = -\mathbfit{n}_B\cdot [\vect{\omega}\times(\vect{\omega}\times\mathbfit{r})],
\eeq 
where $\mathbfit{n}_B$ is the unit vector directed along the magnetic field line, ``$\cdot$" and ``$\times$" denote the scalar and the vector productions respectively.
In the particular case of the orthogonal rotator (i.e., $\alpha=\pi/2$), equation (\ref{eq:a_cent_gen}) is reduced to
\beq
a_{{\rm cen},||}(\alpha=\pi/2) =\omega^2 R_{\rm max}\cos^2\lambda \cos\chi.
\eeq

\begin{figure*}
\centering 
\includegraphics[width=18.cm]{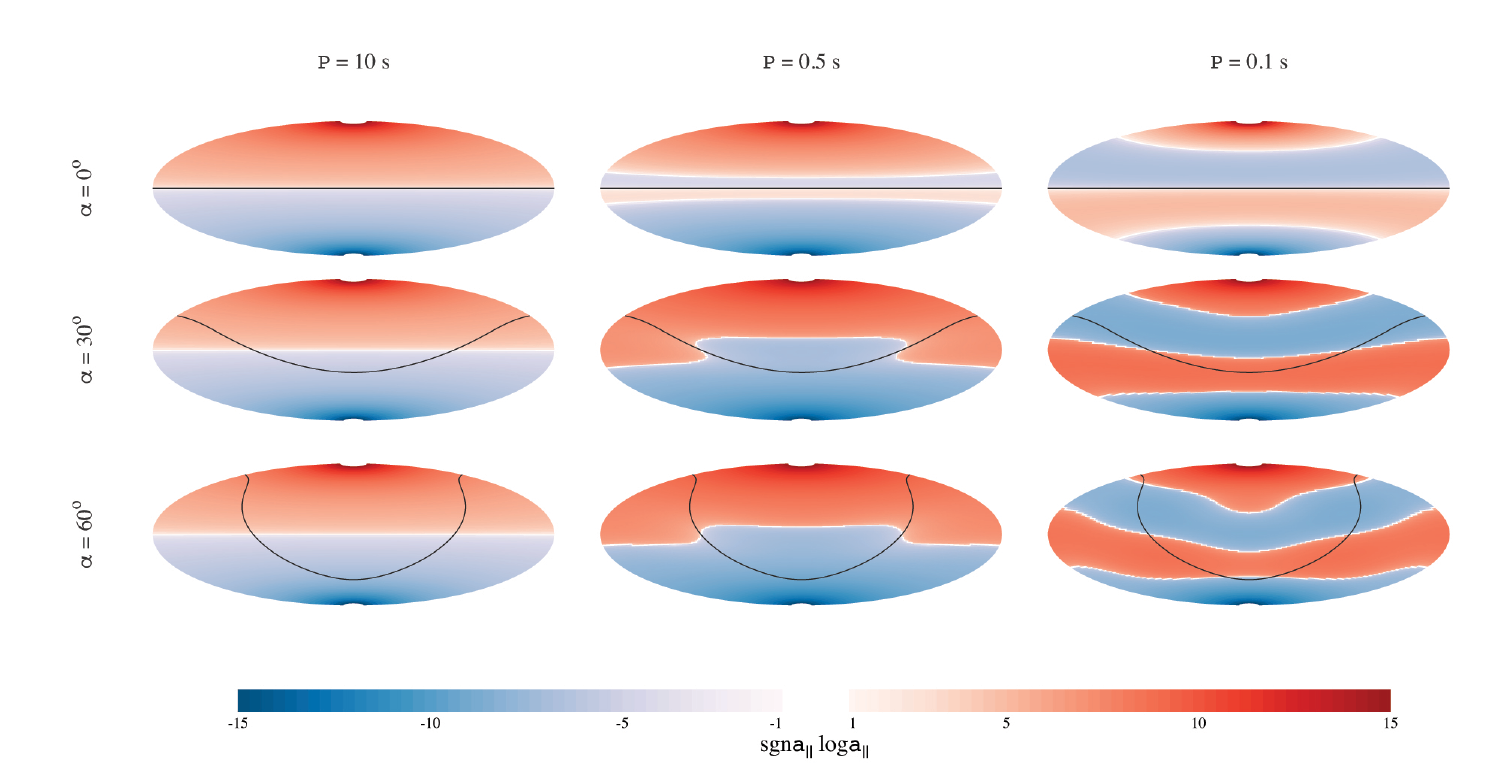}
\caption{
The colour maps represent acceleration along dipole magnetic field lines due to the gravitational and centrifugal forces at the magnetospheric surface.
The red (blue) colour corresponds to the case of the acceleration directed towards the larger (smaller) latitudes, i.e. towards the upper (lower) pole of a star.
The solid black line represents coordinates where the disc plane crosses the dipole surface.
The magnetic obliquity is taken to be $\alpha=0$ (the first line), $30\degr$ (the second line) and $60\degr$ (the third line).
Left, middle and right columns correspond to different NS spin periods: $P_{\rm spin}=10\,{\rm s}$, $0.5\,{\rm s}$ and $0.1\,{\rm s}$ respectively.
Parameters: $m=1.4$, $R_{\rm m}=10^8\,{\rm cm}$.
}
\label{pic:sc_env_acc_03}
\end{figure*}

The total acceleration along the field lines due to the gravitational and centrifugal forces is dependent on the inclination of the magnetic dipole with respect to the rotation axis and NS spin period.
In general, the acceleration is dependent both on the latitude and the azimuth angle (see Fig.\,\ref{pic:sc_env_acc_03}). 
Sufficiently small spin periods result in the appearance of regions with acceleration directed towards the disc plane, i.e. total acceleration due to the gravitational and centrifugal forces tends to prevent accretion flow from its motion towards the poles of a NS. 
It is the centrifugal barrier, which can lead to the propeller regime of accretion in XRPs \citep{1975A&A....39..185I,2006ApJ...646..304U}. 
It is expected that the accretion flow is able to penetrate through the centrifugal barrier in the case of the sufficiently high initial velocity of the flow or in the case of a sufficient geometrical thickness of the disc (some aspects of the influence of accretion disc geometrical thickness on the propeller regime were discussed by \citealt{2022arXiv221112945C}). 
In the case of a magnetic dipole inclined to the rotation axis, the acceleration becomes dependent on the azimuthal coordinate on the dipole surface $\varphi$
(see Fig.\,\ref{pic:sc_env_acc_03}).
At a sufficiently small spin period and sufficiently large inclination of the magnetic dipole with respect to the disc plane, there are regions at the dipole surface where material can stay stably, i.e. the acceleration due to the gravitational and centrifugal forces is directed towards these regions (appearance of these regions was discussed by \citealt{2020MNRAS.496...13A} and \citealt{2023MNRAS.520.4315L}).
The inclination of the magnetic dipole, initial accretion flow velocity, NS spin period and thickness of the disc determine the regions where the material is collected.

\begin{figure*}
\centering 
\includegraphics[width=18.cm]{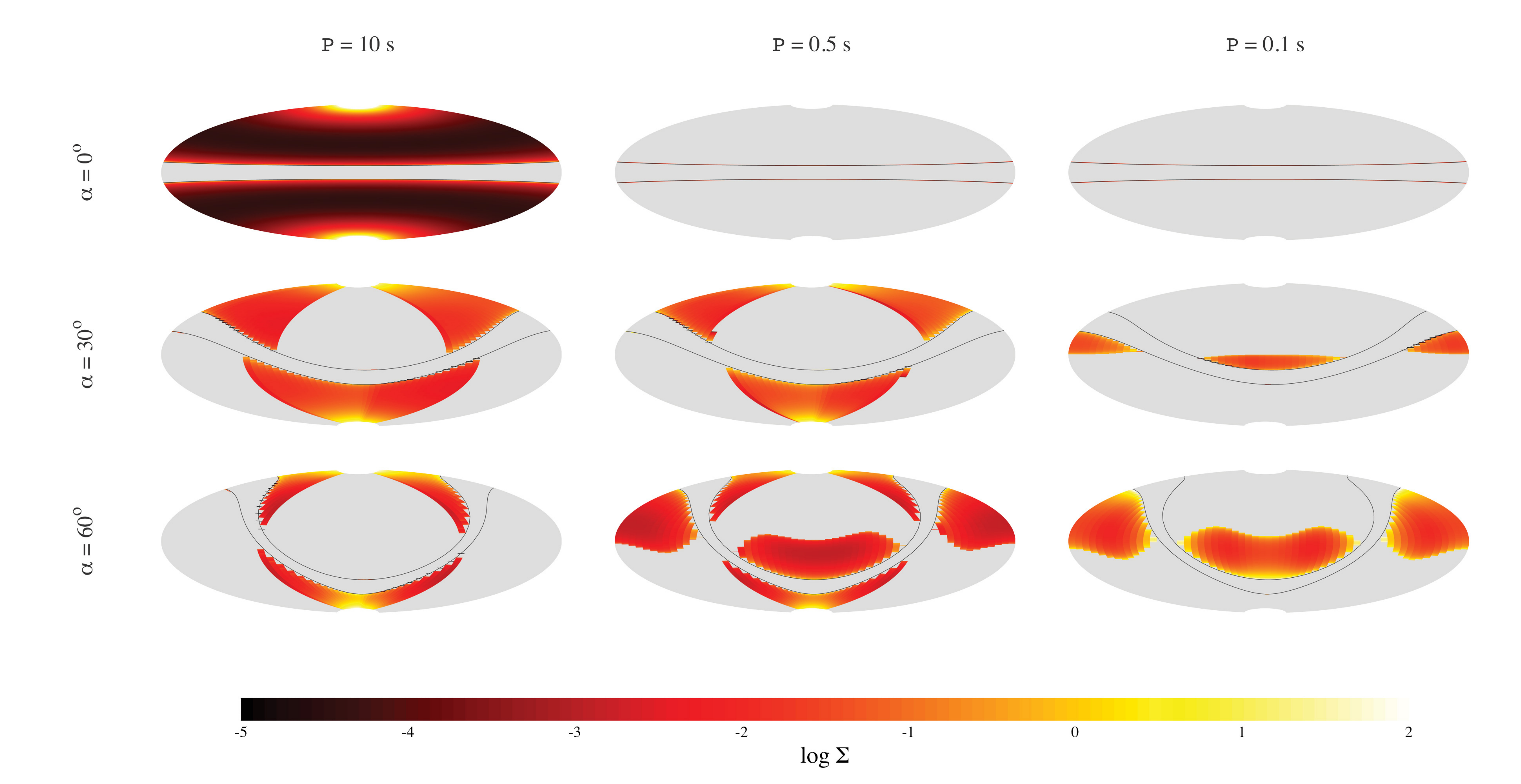}
\caption{
The maps illustrating the distribution of accretion flow surface density over the magnetospheric surface.
Different maps correspond to different NS spin periods ($P=10\,{\rm s}$, $0.5\,{\rm s}$ and $0.1\,{\rm s}$) and various magnetic obliquity $\alpha$. 
The inclination of the magnetic axis with respect to the rotation axis of a NS results in a fractional covering of the magnetosphere.
In panels corresponding to $\alpha=0\degr$ and $P=0.5\,{\rm s}$,
$\alpha=0\degr$ and $P=0.1\,{\rm s}$,
$\alpha=30\degr$ and $P=0.1\,{\rm s}$, and
$\alpha=60\degr$ and $P=0.1\,{\rm s}$
the accretion flow does not reach the NS surface due to the centrifugal force and we see the propeller effect in action. 
In the case of large obliquity, sufficiently fast rotation of a NS results in the appearance of regions where material tends to accumulate but does not move towards a NS.
Parameters: $\dot{M}_{\rm m}=10^{17}\,{\rm g\,s^{-1}}$, $m=1.4$, $R_{\rm m}=10^8\,{\rm cm}$.
}
\label{pic:sc_env_acc_05}
\end{figure*}

Interaction of magnetospheric flow with X-ray photons emitted by the central objects results in appearance of radiative force. 
The radiative force is determined by local X-ray energy flux at the magnetospheric surface, the fraction of intercepted radiation and local inclination of magnetospheric surface in respect to the predominant direction of photons momentum.
In the first approximation, the radiative force is directed in oppositely to the gravitational force.
The fraction of intercepted X-ray photons is determined by the surface density of magnetospheric flow and mechanism of opacity.
In our simulations, we assume that the opacity is dominated by the non-magnetic Compton scattering and taken to be $\kappa_{\rm e}=0.34\,{\rm cm^2\,g^{-1}}$.
Non-magnetic opacity is a valid approximation because magnetic field strength decreases rapidly with a distance from a NS and the cyclotron energy $E_{\rm cyc}\simeq 11.6\,B_{12}\,{\rm keV}$ on the most of magnetospheric surface is much smaller then the typical energy of photons.
In the limiting case of total absorption of radiation, the radiative force along the dipole field lines acting on a unit area can be estimated as
\beq\label{eq:f_r_app1}
\frac{\d f_{\rm R}}{\d S}(\lambda)=\sin\chi \cos\chi\, \frac{F(\lambda)}{c}, 
\eeq
where $F(\lambda)$ is the photon energy flux.
In the real situation, only a fraction of photons are absorbed/scattered by the envelope and contribute to be radiative force.
Thus, (\ref{eq:f_r_app1}) has to be rewritten as
\beq\label{eq:f_r_app2}
\frac{\d f_{\rm R}}{\d S}(\lambda)=\sin\chi \cos\chi\, \frac{F(\lambda)}{c}\left(1 - e^{-\Sigma \kappa_{\rm e}/\sin\chi}\right), 
\eeq 
where $\Sigma$ is the local surface density of the magnetospheric flow.
The estimation (\ref{eq:f_r_app2}) assumes that the scattering is isotropic and, therefore, the photons effectively transfer their momentum to the accretion flow due to the first scattering event only. 
{
The equations (\ref{eq:f_r_app1}) and (\ref{eq:f_r_app2}) can be rewritten via the radiation force acceleration to keep the uniform with the previous description of the forces. 
The radiation acceleration inside the accretion flow is
\beq 
  a_{\rm rad}(\lambda) = \cos\chi\, \frac{\kappa_{\rm e}}{c}\,F(\lambda)\,e^{-m \kappa_{\rm e}/\sin\chi},
\eeq 
where $m$ is Lagrangian coordinate of the point inside the flow. 
It is determined as 
$\d m = \rho\,\d z$, 
where $z$ is a coordinate along the local normal to the magnetospheric surface, and $\rho$ is the plasma mass density.
The radiation acceleration averaged across the accretion flow is
\beq\label{eq:a_rad_ave}
  a_{\rm rad}(\lambda,\Sigma) =  \sin\chi \cos\chi\, \frac{F(\lambda)
  }{c\,\Sigma}\left(1-e^{-\Sigma \kappa_{\rm e}/\sin\chi}\right),
\eeq  
where $\Sigma$ is a total surface density integrated across the accretion flow at the given position. The last equation coinsides with (\ref{eq:f_r_app2}), if we take that $a_{\rm rad}(\lambda,\Sigma)\,\Sigma = \d f_{\rm R}/\d S$.}
In average, the photon scatterings following the first one do not affect the dynamics of the flow. 

In the case of optically thick flow (i.e., $\Sigma\kappa_{\rm e}>1$), the photon energy flux varies over the optical depth: some photons are scattered and reflected back by the surface layers, and only a fraction of photons penetrates deeper inside the flow covering magnetic dipole.
Therefore, the acceleration due to the radiative force is dependent on the optical depth, and the layers located closer to the central source experience stronger accelerations.
It might result in a gradient of velocity across the accretion channel. 
In this paper, we do not account for the influence of the radiative force gradient and possible effects of viscosity inside the accretion channel.

\subsection{Radiation from the neutron star surface}

NS radius is expected to be much smaller than its magnetosphere (see Eq.\,\ref{eq:Rm}), which is a good approximation in the case of XRPs. 
Thus, accounting for the influence of radiative force on the accretion flow dynamics, we consider a NS as a point source of X-ray radiation.
Because accretion flow is directed towards NS magnetic poles by the stellar magnetic field, the emission from the central object can be strongly anisotropic. 
The angular distribution of X-ray luminosity is not known and described in our model by a special parameter $a_{\rm sp}\in [-1;+1]$ in our model: 
\beq \label{eq:beam}
\frac{\d L_i}{\d\Omega}\propto 1 + a_{\rm sp} |\cos\theta_{B,k}|,
\eeq 
where index $k\in\{1,2\}$ denotes the pole of a NS, and
$\theta_{B,i}\in[0,\pi]$ is the colatitude in the reference frame related to the magnetic dipole.
Parameter $a_{\rm sp}=0$ corresponds to the case of isotropic radiation from the central object,
$a_{\rm sp}=-1$ corresponds to the fan beam diagram, 
and $a_{\rm sp}>0$ corresponds to pencil beam diagram.
The beam pattern is shaped by the geometry of the emitting region at the NS surface \citep{1973A&A....25..233G} and can be strongly influenced by the gravitational bending in the vicinity of a NS \citep{1988ApJ...325..207R,2001ApJ...563..289K,2018MNRAS.474.5425M}.

The total luminosity of the central object 
$L_{\rm tot}=L_1+L_2$ 
can be variable and dependent on the current mass accretion rate onto the poles of a NS.
The mass accretion rate on one pole can be different from the mass accretion rate on another pole, which results in a difference between $L_1$ and $L_2$ and the corresponding X-ray energy flux at the magnetospheric surface: 
\beq\label{eq:L2F}
F (\theta,t) = \frac{1}{r^2}\frac{\d L_{\rm tot}(t)}{\d\Omega}, 
\eeq 
where $r$ is determined by (\ref{eq:dip_fl}).

\subsection{Mass accretion rate at the inner disc radius}

The average mass accretion rate at the inner disc radius is a parameter of our simulations.
It can be considered to be constant or variable due to fluctuating mass accretion rate in the disc (see, e.g., \citealt{2019MNRAS.486.4061M}).
{
Stochastic variability of the mass accretion rate at the inner disc radius is described by the power density spectra (PDS) and root-mean square (rms) variability.
The broad-band component of the PDS in X-ray binaries is generally well described by a twice-broken power law \citep{1993ApJ...411L..79H}.
In this paper, we restrict ourselves to investigating the effect of the rms on the magnetospheric accretion and, for simplicity, assume that the PDS is given by a power law in a given range of Fourier frequencies $f$, i.e. 
\beq\label{eq:PDS}
{\rm PDS}\propto f^q,
\eeq
where $q$ is fixed at $(-1)$.
}
Because the disc can be geometrically thin or thick depending on the mass accretion rate (see estimations of the disc thickness in \citealt{2007ARep...51..549S}), the particles participating in our simulations start their motion along dipole magnetic field lines at $\lambda_0\ne \lambda_{\rm d}=\pi/2 - \theta_{\rm d}$.

\section{Numerical model}
\label{sec:NumMod}

We consider a one-dimentional motion of accretion flow along dipole field lines given by (\ref{eq:dip_fl}).
In the case of the rotational axis orthogonal to the disc plane (i.e. $\alpha=0$), the accretion flow starts its motion at the latitude 
\beq 
\lambda_0 = \frac{\pi}{2}-\theta_{\rm d}\pm \delta\lambda,
\eeq 
where $\delta\lambda$ is determined by the semi-thickness of the accretion disc at the magnetospheric radius and reaches the NS surface at coordinate 
\beq
\lambda_{\rm NS}\simeq{\rm acos}(\sqrt{R_{\rm NS}/R_{\rm m}}).
\eeq
The accretion channel in between coordinates $\lambda_0$ and $\lambda_{\rm NS}$ is divided into 
$N_{\lambda}$ intervals in $\lambda$-coordinate 
and 
$N_{\varphi}$ intervals in the azimuthal angle.
As a result, we construct a two-dimensional grid composed of $N_{\lambda}\times N_\varphi$ cells covering the magnetospheric surface.
For the simulations represented in this paper, we have taken $N_\lambda=630$ and $N_\varphi = 24$.
Each simulation covers time interval of $30\,{\rm s}$, is composed of a series of time steps and traces the motion of quasi-particles along the field lines.

At each time step of the simulation, we have information about the coordinates of all particles, their velocities and the total mass within each cell of a grid.
At each time step, we perform the following procedures:
\begin{enumerate}[leftmargin=*]
\item 
We start by determining the mass accretion rate at the inner disc radius based on the average mass accretion rate and rms of mass accretion rate variability.
The mean mass accretion rate and its rms are parameters in our simulations.
We follow the algorithm proposed by \cite{1995A&A...300..707T} to get a time series of the mass accretion rate and thus the mass accretion rate at a given moment.
In our calculations, we take $q=-1$ and simulate the mass accretion rate time series on the base of assumed rms and PDS in the Fourier frequency interval $0.1\,{\rm Hz}<f<100\,{\rm Hz}$.
To model the time series, we take $5\times 10^3$ frequencies in the considered frequency range \citep[for details see][]{1995A&A...300..707T}.
\item 
\red{We add $N_{\rm p}$ particles of equal mass to each of $N_{\varphi}$ grid cells located near the edge of accretion disc,} i.e. we add $N_{\rm p}N_{\varphi}$ particles.
The mass of particles added to the grid at a given time step $t_i$ is related to the mass accretion rate $\dot{M}$ at the inner disc radius as
\beq
m_{{\rm p},j}=\frac{\dot{M}(t_i)\Delta t_i}{N_{\rm p} N_\varphi},
\eeq
where $\Delta t_i$ is a time interval of the simulation and index $j$ enumerates particles participating in a simulation.
\red{In the simulations represented in this paper, we use $N_{\varphi}=24$ and $N_{\rm p}=2$.}
The particles are injected to the accretion flow at a random position within the cell located at the accretion disc plane, i.e. the initial coordinates are
\beq
\lambda_j^{\rm (ini)} = \lambda_0 + X_j\,\frac{\lambda_{\rm NS}-\lambda_0}{N_\lambda}, 
\eeq
where $X_j\in (0;1)$ is a random number.
The initial velocity of each particle along magnetic field lines $v_{\rm ini}$ is a parameter of the simulation.
It is expected to be close to the thermal velocity of protons at the inner disc radius:
\beq 
v_{\rm ini}\sim v_{\rm p}\approx 
3\times 10^7\,T_{\rm keV}^{1/2}\,{\rm cm\,s^{-1}},
\eeq 
where $T_{\rm keV}$ is the plasma temperature at $R_{\rm m}$ in units of keV.
\red{
In our simulations we use $v_{\rm ini}\sim 10^7\,{\rm cm\,s^{-1}}$.}
\item \label{get_Sigma}
On the base of known coordinates of particles and their masses, we get the total mass in each cell of the grid \red{and local surface density of the magnetospheric accretion flow $\Sigma(\lambda,\varphi,t_i)$.}
\item
We calculate the time interval towards the new time step.
The time step of the simulation is variable and dependent on the typical particle velocities in the cells. 
It is taken to be sufficiently small to prevent particle transitions to the cells further than the nearest ones
\beq\label{eq:Delta_t}
\Delta t_i=\min\left\{
\min\limits_{j}\left[\frac{1}{5}\frac{(\d x/\d\lambda) \Delta\lambda_j}{v_j(t_i)}\right],
10^{-5}\,{\rm s}
\right\},
\eeq
where $v_j$ is a velocity of a particle $j$, and the minimum is taken over all particles currently participating in simulation.
\item 
We get the acceleration of particles along magnetic field lines. 
The particle motion is calculated by accounting for gravitational, radiation and centrifugal forces (see Section\,\ref{sec:Dynamics}).
We calculate the vector sum of three forces for each cell of the grid and get acceleration along field lines for each particle currently participating in the simulation.
The gravitational and centrifugal accelerations are not dependent on time and can be pre-calculated.
The radiative force, on the contrary, can be variable due the variations of accretion luminosity and local surface density of accretion flow covering NS magnetosphere. 
\item 
On the basis of calculated accelerations for particles, we get their velocities
\beq 
v_{j}(t_{i+1})=v_{j}(t_{i})+ a_j(t_i)\Delta t_i
\eeq   
and coordinates 
\beq 
\lambda_{j}(t_{i+1}) &=& \lambda_j(t_i) + \Delta\lambda_j(t_i), \\
\Delta\lambda_j(t_i) &=& \frac{\Delta x_j(t_i)}
{R_{\rm m}\cos\lambda_j(t_i)[1+3\sin^2\lambda_j(t_i)]^{1/2}},
\nonumber \\
\Delta x_j(t_i) &=& \frac{ v_j(t_{i+1})+v_j(t_{i}) }{2} \Delta t_i.\nonumber
\eeq
\item  
We update the list of particles participating in the simulation. 
In particular, we account for particles accreted to the NS magnetic poles: particles that have reached one of the magnetic poles are taken into account when calculating the mass accretion rate and are removed from the list of particles.
The mass accretion rate onto a NS magnetic pole is calculaterd as
\beq\label{eq:dot_m}  
\dot{M}_k (t_i)= \frac{\sum_{j} m^{(k)}_{{\rm p},j}}{\Delta t_i}, 
\eeq  
where $k\in \{1,2\}$ denotes one of the magnetic poles and the summation is performed over the particles that have reached the surface in the time interval $\Delta t_i$.
We also account for particles that have returned to the accretion disc. 
These particles remain in the simulation but are restarted from a random azimuthal coordinate at the next time step.
The mass accretion rate calculated in (\ref{eq:dot_m}) and final velocity of particles are used to get accretion luminosity
\beq\label{eq:luminosity}  
L_k (t_i)= \frac{\sum_{j} (\gamma_j -1)m^{(k)}_{{\rm p},j}c^2}{\Delta t_i}, 
\eeq   
where $\gamma_j=[1-(v_j/c)^2]^{-1/2}$ is the final Lorentz factor of a particle.
The luminosity (\ref{eq:luminosity}) is used to calculate photon energy flux (\ref{eq:L2F}) at the magnetospheric surface and then the radiative force at the next time step of simulation.
\item   
We return to step (i).
\end{enumerate}

Particle motion changes the mass density in each grid cell, and, as a result, the influence of three forces on the particles in the cells is recalculated at each time step.

\red{
To investigate features of the mass accretion rate variability in the frequency domain, we redact the time series obtained by (\ref{eq:dot_m}), where the time intervals between element of a time series are variable and short (see equation \ref{eq:Delta_t}), which results in a Poisson noise.
In particular, we combine nearby elements of the simulated time series into one and get an updated time series with larger and homogeneous time $\Delta t\simeq 2\times 10^{-3}\,{\rm s}$.
Using the updated time series of the mass accretion rate at the inner disc radius and at the NS surface, we calculate its Fourier transform:
\beq\label{eq:FT_Mdot}
\overline{\dot{M}}(f) = \Delta t \sum\limits_{j=1}^{N_t} 
\dot{M}_k (t_j)e^{2\pi i f t_j},
\eeq 
where $f$ is the Fourier frequency and $N_t$ is a number of simulated points in the time series. 
The power density spectrum is calculated using 
normalisation:
\beq \label{eq:PDS}
{\rm PDS}=
\frac{2}{M_{\rm acc}}\left|\overline{\dot{M}}(f)\right|^2
\eeq 
where $M_{\rm acc}=\sum_{j=1}^{N_t}\dot{M}_k (t_j)\Delta t$ is the total mass accreted onto a NS during the time interval covered by a time series.
Calculating the Fourier transform (\ref{eq:FT_Mdot}) and PDS (\ref{eq:PDS}) we exclude the first second of the simulation which is required for the accretion flow to reach the NS surface.
The Nyquist frequency for the simulated time series is $f_{\rm N}=0.5/\Delta t\simeq 250\,{\rm Hz}$ and further we discuss PDS shape below this frequency only.
In the case of normalisation given by (\ref{eq:PDS}), the noise caused by finite number of particles participating in a simulation is expected at the level 
\beq 
{\rm PDS}_{\rm noise} \approx
\frac{2\,M_{\rm acc}}{N_{\rm tot}},
\eeq
where $N_{\rm tot}$ is a total number of particles accreted onto the stellar surface during the entire simulation. 
}

\red{
To investigate stability of magnetospheric coverage by the accretion flow, we calculate the local average surface density, local averaged squared surface density 
\beq\label{eq:Sigma_ave}
\left[\begin{array}{c} 
\langle\Sigma(\lambda,\varphi)\rangle \\ 
\langle\Sigma^2(\lambda,\varphi)\rangle \end {array}\right]
 = 
\frac{1}{t_{\rm tot}}\sum\limits_{i}
\Delta t_i
\left[\begin{array}{c} 
\Sigma(\lambda,\varphi,t_i) \\ 
\Sigma^2(\lambda,\varphi,t_i)\end {array}\right],
\eeq 
and its local standard deviation
\beq\label{eq:sigma_Sigma}
\sigma_\Sigma(\lambda,\varphi) = 
\sqrt{
\langle\Sigma^2(\lambda,\varphi)\rangle -
\langle\Sigma(\lambda,\varphi)\rangle^2
}.
\eeq 
}

\section{Results of numerical simulations}
\label{sec:NumRes}

\begin{figure*}
\centering 
\includegraphics[width=18.cm]{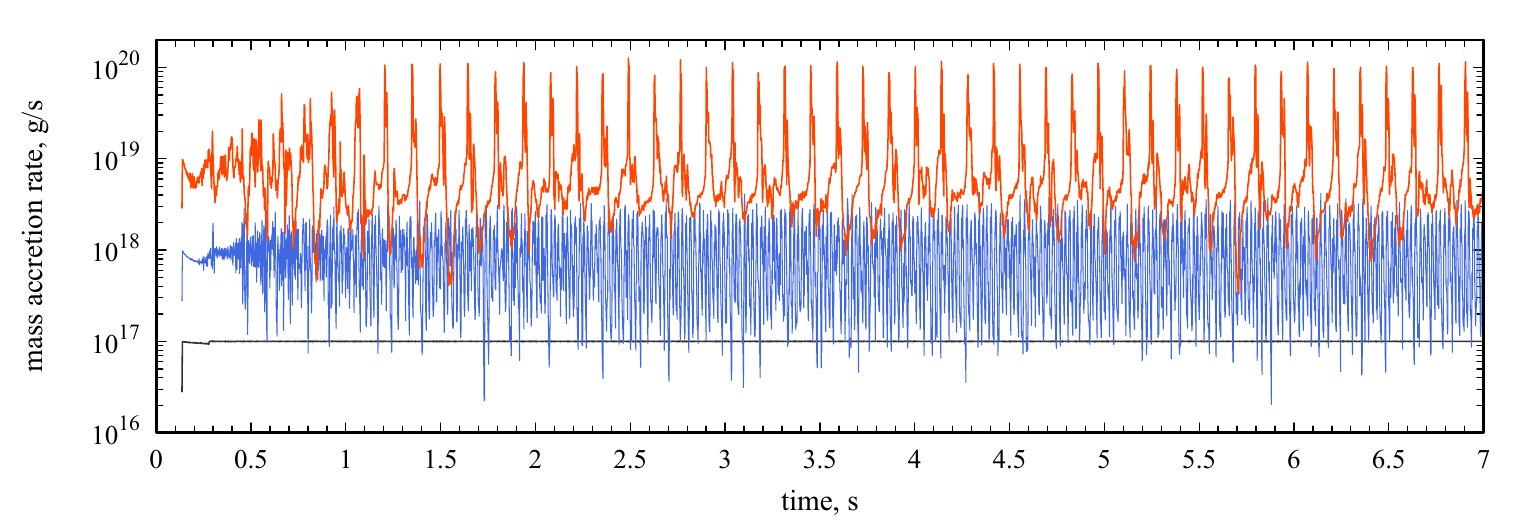} 
\caption{
The mass accretion rate onto the NS surface calculated for different constant mass accretion rates at the inner disc radius: 
$\dot{M}(R_{\rm m})=10^{17}\,{\rm g\,s^{-1}}$ ($\sim 0.1\dot{M}_{\rm Edd}$, black line), 
$10^{18}\,{\rm g\,s^{-1}}$ ($\sim \dot{M}_{\rm Edd}$, blue line), and $10^{19}\,{\rm g\,s^{-1}}$ ($\sim 10\dot{M}_{\rm Edd}$, red line).
In the case of low mass accretion rates (black line), the mass accretion rate at the NS surface replicates the mass accretion rate at the inner disc radius.
Sufficiently high mass accretion rate from the disc results in appearance of instability of accretion process over the magnetospheric surface, which makes mass accretion rate at the NS surface strongly variable (blue and red lines).
At the mass accretion rates $\sim 10^{19}\,{\rm g\,s^{-1}}$ from the disc (red line), the mass accretion rate at the stellar surface shows quasi-periodic oscillations.
Parameters: $m=1.4$, 
$R_{\rm m}=10^{8}\,{\rm cm}$, $P_{\rm spin}=10\,{\rm s}$, isotropic central source, and $\alpha=0$.
}
\label{pic:sc_env_dyn_1}
\end{figure*}

\begin{figure}
\centering 
\includegraphics[width=8.7cm]{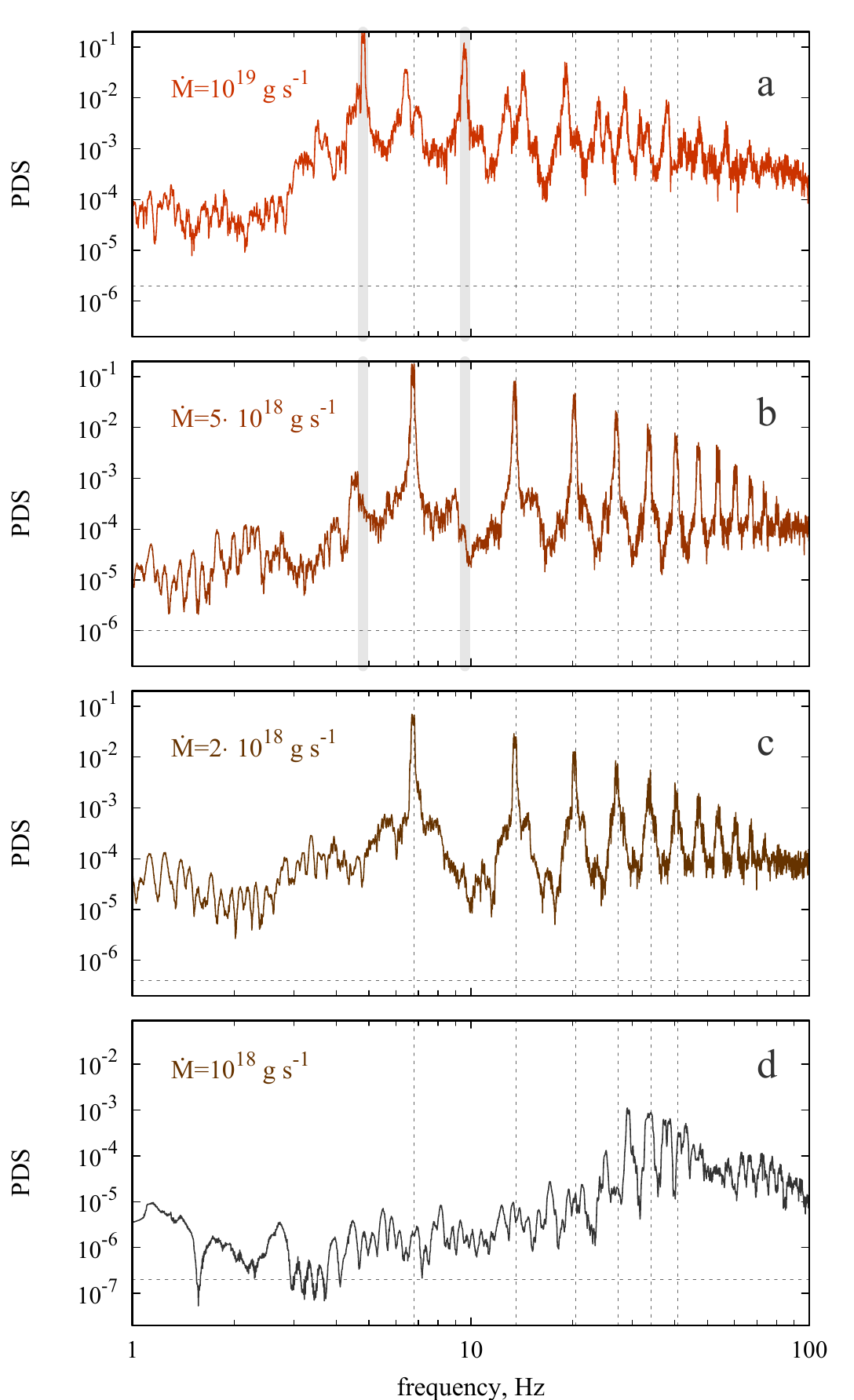} 
\caption{
\red{
The modeled PDS of the mass accretion rate fluctuations at the NS surface.
Different panels show the PDS for different mass accretion rates from the disc: 
$10^{19}\,{\rm g\,s^{-1}}$,
$5\times 10^{18}\,{\rm g\,s^{-1}}$,
$2\times 10^{18}\,{\rm g\,s^{-1}}$, and
$10^{18}\,{\rm g\,s^{-1}}$ (from top to bottom).
Parameters: $R_{\rm m}=10^8\,{\rm cm}$, $P_{\rm spin}=10\,{\rm s}$, isotropic source of emission.
} 
}
\label{pic:sc_env_dyn_6}
\end{figure}

\begin{figure*}
\centering 
\includegraphics[width=18.cm]{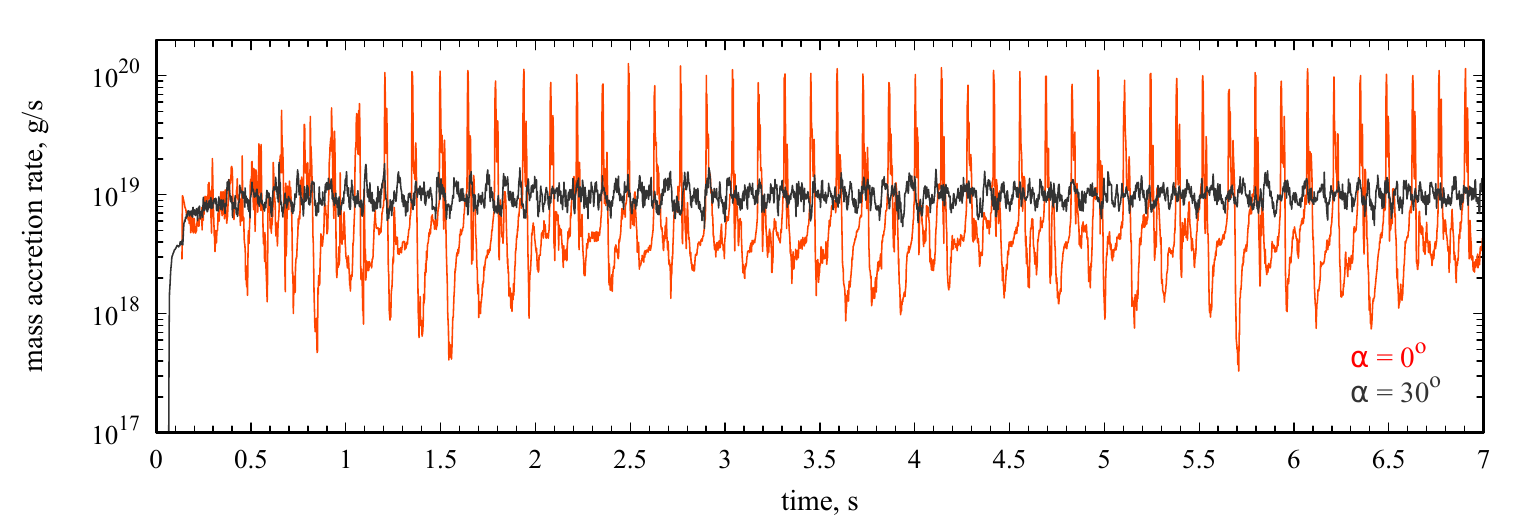} 
\caption{
The mass accretion rate onto NS surface calculated for different inclinations of magnetic dipole in respect to the accretion disc plane: 
$\alpha=0$ (red line) 
and 
$\alpha = 30^\circ$ (black line).
The mass accretion rate at the inner disc radius is constant and fixed at $\dot{M}=10^{19}\,{\rm g\,s^{-1}}$ ($\sim 10\dot{M}_{\rm Edd}$).
Parameters: $m=1.4$, 
$R_{\rm m}=10^{8}\,{\rm cm}$, 
$P_{\rm spin}=10\,{\rm s}$, isotropic central source.
}
\label{pic:sc_env_dyn_2}
\end{figure*}

\begin{figure}
\centering 
\includegraphics[width=8.7cm]{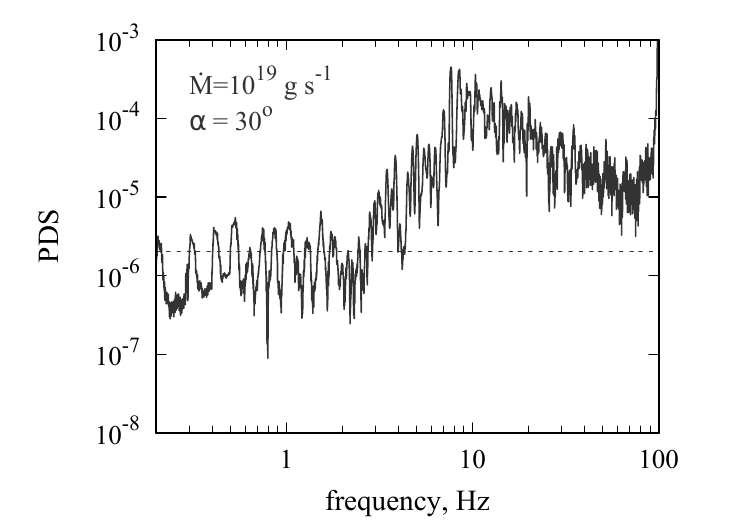} 
\caption{
\red{
The modeled PDS of the mass accretion rate fluctuations onto the NS surface in a system, where a NS is inclined with respect to the disc plane.
We do not see QPO peaks, but there is a broadband component appear at high frequencies ($\gtrsim 10\,{\rm Hz}$). 
Horizontal dashed line illustrates the level of Poisson noise in the simulation. 
Parameters: $\dot{M}=10^{19}\,{\rm g\,s^{-1}}$, $R_{\rm m}=10^8\,{\rm cm}$.} 
}
\label{pic:sc_PDS_02}
\end{figure}

\begin{figure*}
\centering 
\includegraphics[width=18.cm]{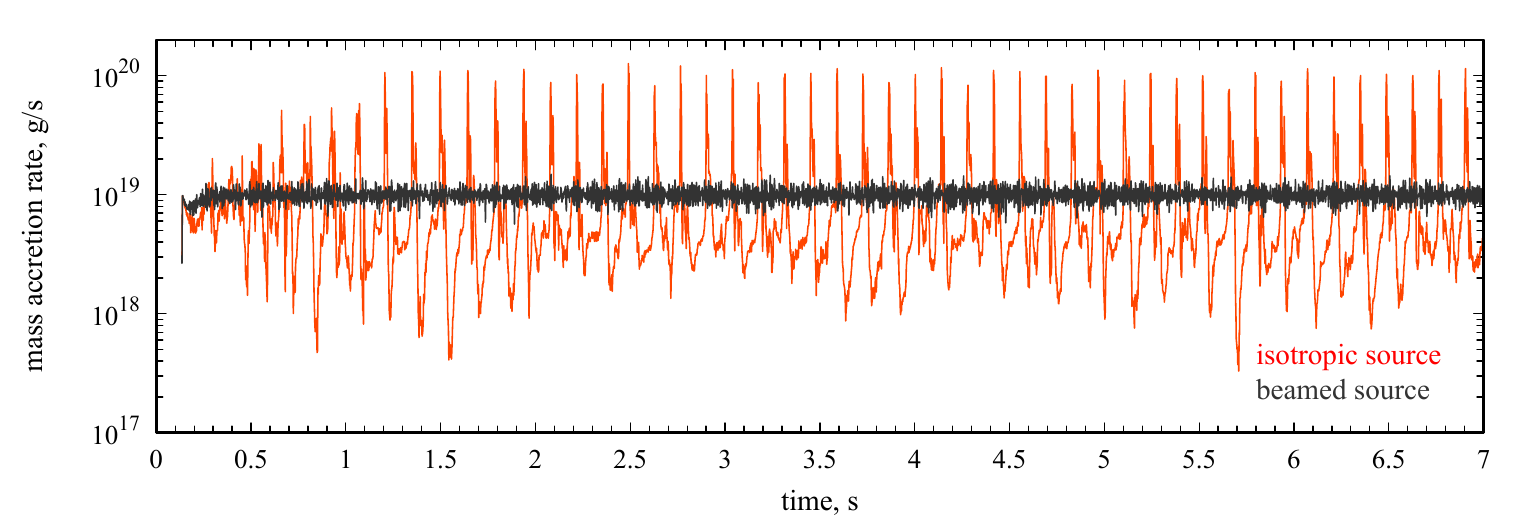} 
\caption{
The mass accretion rate onto NS surface calculated for different beam patterns: the red line illustrates the case of isotropic central source, while the black line corresponds to the case of beamed emission from the poles of a NS (parameter $a_{\rm sp}=2$ in \ref{eq:beam}).
The mass accretion rate at the inner disc radius is constant and fixed at $\dot{M}=10^{19}\,{\rm g\,s^{-1}}$.
The rotation axis is taken to be orthogonal to the accretion disc plane. 
Parameters: $m=1.4$, 
$R_{\rm m}=10^{8}\,{\rm cm}$, $P_{\rm spin}=10\,{\rm s}$.
}
\label{pic:sc_env_dyn_3}
\end{figure*}

\begin{figure}
\centering 
\includegraphics[width=8.5cm]{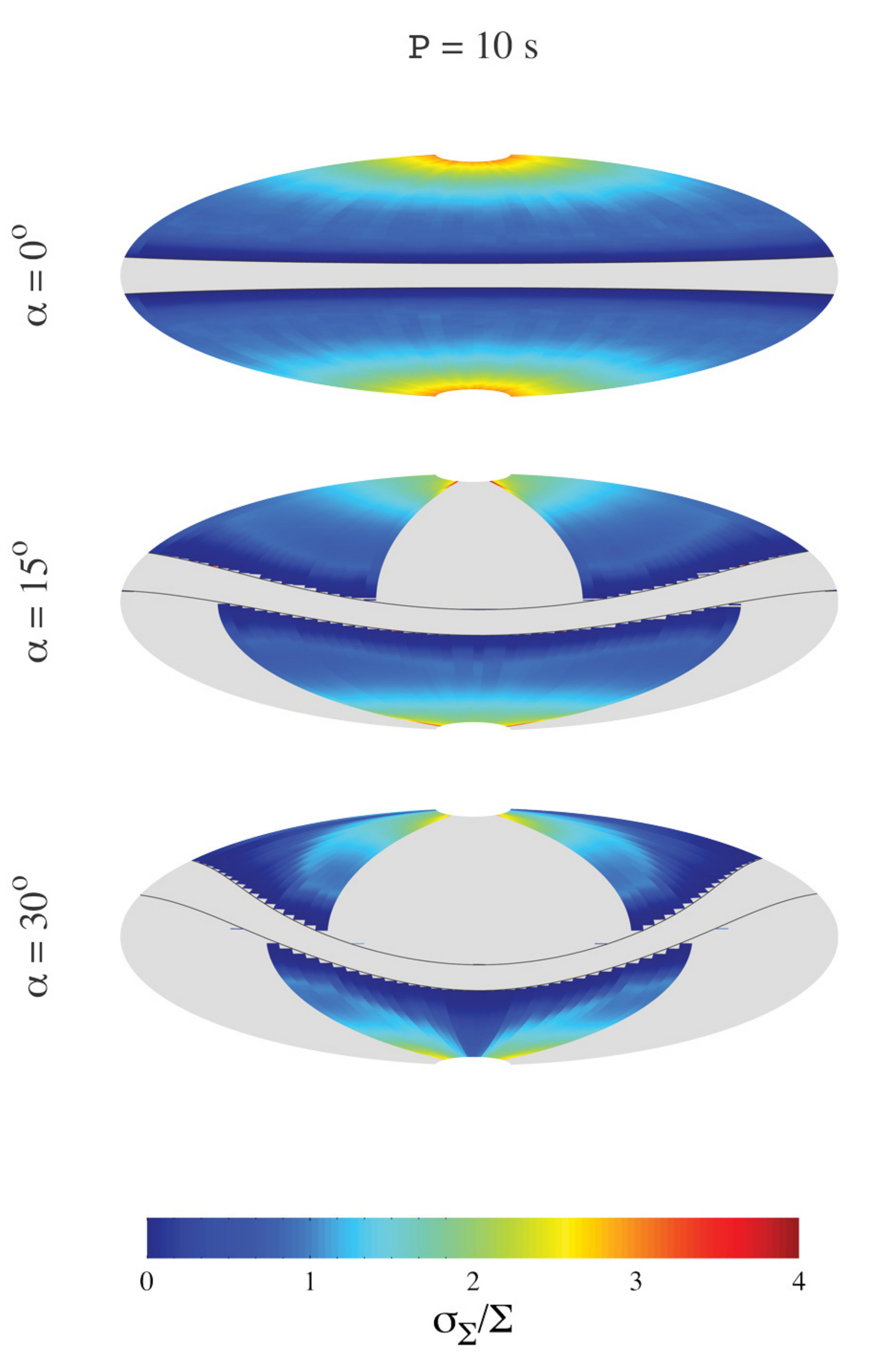}
\caption{
Maps of the relative standard deviation of the surface density \red{(see equaions \ref{eq:Sigma_ave} and \ref{eq:sigma_Sigma})} on the magnetospheric surface for the mass accretion rate fixed at $\dot{M}_{\rm m}=10^{19}\,{\rm g\,s^{-1}}$ and different magnetic obliquity $\alpha=0\degr,\,15\degr$ and $30\degr$ (from top to bottom).
In the case of aligned rotation and magnetic axis, the relative variability of the surface density is stronger at larger latitudes.
In the case of inclined magnetic dipole, only a fraction of magnetospheric surface is covered by accretion flow and the variability becomes dependent on the azimuthal angle: the regions located closer to the edges of accretion curtain tend to be more variable.
Parameters: 
$m=1.4$, 
$R_{\rm m}=10^8\,{\rm cm}$, 
$P=10\,{\rm s}$.
}
\label{pic:sc_env_acc_06}
\end{figure}

In our simulations of accretion flow dynamics, we initialize the flow at $t=0$ from the edges of the accretion disc, which is fixed radius $R_{\rm m}=10^8\,{\rm cm}$.
It takes a fraction of a second (approximately $0.1\,{\rm s}$ for the specified $R_{\rm m}$) for the accretion flow to reach the NS surface. 
This duration is slightly longer than the free-fall time from the inner disc radius due to the flow following curved magnetic field lines. 
The radiative force comes into play as soon as the flow reaches the stellar surface.
Before that, the flow dynamic is shaped by the gravitational and centrifugal forces only.
In this section, we first discuss the results of simulations performed for a low mass accretion rates from the disc ($\dot{M}\lesssim 10^{17}\,{\rm g\,s^{-1}}$), when  the influence of the radiative force can be neglected (Section\,\ref{sec:res_grav_and_cent}).
In the following we discuss the effects introduced by strong radiative force at high accretion rates ($\dot{M}\gtrsim \dot{M}_{\rm Edd} \sim 10^{18}\,{\rm g\,s^{-1}}$, Section\,\ref{sec:res_rad}).

\subsection{The influence of gravitational and centrifugal forces}
\label{sec:res_grav_and_cent}

The accelerations along $B$-field lines due to the NS gravity and the centrifugal force are dependent on the mass of a NS, spin period, size and inclination of the magnetic dipole with respect to the disc plane.
The maps of this acceleration are pre-calculated (see Fig.\,\ref{pic:sc_env_acc_03}).
Sufficiently fast rotation of a NS magnetosphere results naturally in arising of the centrifugal barrier, i.e. the regions at the magnetospheric surface, where the local acceleration is directed back to the disc plane. 
The centrifugal barrier is stronger in the case of faster rotation of a NS.
It is expected that the material from the disc can penetrate through the centrifugal barrier in the case of the sufficiently high initial velocity of the material (in this case, the barrier slows down the flow but cannot stop it) or sufficiently large geometrical thickness of accretion disc, when the matter is collected from the upper layers of the disc (the latest effect was discussed by \citealt{2022arXiv221112945C}). 

In the case of an inclined magnetic dipole, there are regions at the inner disc radius where the centrifugal and gravitational forces prevent material motion towards one of the magnetic poles of a NS. 
It results in a partial covering of NS magnetosphere by the flow from the disc (see Fig.\,\ref{pic:sc_env_acc_05}).
Additionally, in the case of the sufficiently fast rotation of inclined magnetic dipole, there are regions at the NS magnetosphere where the projections of gravitational and centrifugal forces cancel each other.
The appearance of these regions of stable equilibrium may result in local accumulation of accreting material (see Fig.\,\ref{pic:sc_env_acc_05}). 
The fate of the matter locally accumulated at the magnetospheric surface might be affected by the cooling and heating processes.
In particular, sufficiently fast cooling of matter accumulated at the magnetosphere can result in events of episodic accretion \citep{2012MNRAS.420..216S}.
Analyses of heating and cooling processes, however, require specific analyses, which are out of the scope of this paper. 

According to our simulations, the coverage of the NS magnetosphere is dependent both on the inclination of the magnetic dipole and NS spin period: the larger the inclination and the faster the rotation of a NS, the smaller the area of magnetosphere covered by accretion flow (see Fig.\,\ref{pic:sc_env_acc_05}). 
We note that the arising of the ``propeller" effect in our simulations is dependent on the inclination of magnetic dipole: the spin period of $0.5\,{\rm s}$ is small enough to stop accretion in the case of $\alpha=0\degr$, but still allows accretion at $\alpha=30\degr$ and $\alpha=60\degr$  (see Fig.\,\ref{pic:sc_env_acc_05}).

\subsection{Radiative force coming into play}
\label{sec:res_rad}

Influence of the radiative force depends on the mass accretion rate and corresponding energy release at the NS surface.
At low, sub-Eddington ($\dot{M}\ll 10^{18}\,{\rm g\,s^{-1}}$), mass accretion rates, the radiative force does not play a significant role and stable mass accretion rate at the inner disc radius is replicated (with a time delay) at the stellar surface (see black line corresponding to $\dot{M}=10^{17}\,{\rm g\,s^{-1}}$ in Fig.\,\ref{pic:sc_env_dyn_1}). 
High mass accretion rates, however, result in high luminosity and radiative force comparable to the gravitational and centrifugal forces. 
{It causes instability of accretion flow at the mass accretion rates above the Eddington value ($\dot{M}\gtrsim 10^{18}\,{\rm g\,s^{-1}}$, see blue and red curves in Fig.\,\ref{pic:sc_env_dyn_1} corresponding to the mass accretion rates $10^{18}\,{\rm g\,s^{-1}}$ and $10^{19}\,{\rm g\,s^{-1}}$ from the disc respectively) and sharp increase in rms of the mass accretion rate variability at the NS surface. 
Sufficiently intensive mass inflow at the inner disc radius ($\dot{M}\gtrsim 10^{19}\,{\rm g\,s^{-1}}$)} leads to quasi-periodic oscillations (QPOs) of the mass accretion rate at the stellar surface (see red line in Fig.\,\ref{pic:sc_env_dyn_1}).
The time scale of the QPOs is close to the free-fall time scale:
\beq \label{eq:P_QPO}
\red{
P_{\rm QPO} \approx \pi\left(\frac{R_{\rm m}^3}{2 GM}\right)^{1/2}
\simeq 0.2 \left(\frac{R_{\rm m,8}^3}{m}\right)^{1/2}\,{\rm s}.}
\eeq
The mechanism standing behind the QPOs is related to the influence of radiative force on the accretion flow and can be explained as follows: 
(i) accretion flow reaches NS surface and results in high luminosity of the central object;
(ii) high luminosity leads to the radiative force, which is large enough to stop the flow on its way towards a star,
(iii) mass accretion rate onto a NS drops as well as the accretion luminosity, which leads to a drop of the radiative force affecting the magnetospheic flow motion, 
(iv) under condition of small radiative force, accretion flow continues its motion towards a NS and reaches stellar surface within the free-fall timescale,
(v) the cycle returns to the first step and continues quasi-periodically.
\red{The mass accretion rate onto one magnetic pole can differ from the mass accretion rate onto the other one. 
However, there is still a correlation between the two mass accretion rates.
}

\red{
In a high luminosity state, the PDS of the simulated time series of the mass accretion rate at the NS surface shows QPO peaks and its harmonics (see Fig.\,\ref{pic:sc_env_dyn_6}abc).
In the range of mass accretion rate $2\times 10^{18}\,{\rm g\,s^{-1}}\lesssim \dot{M}\lesssim 5\times 10^{18}\,{\rm g\,s^{-1}}$, QPO appears at a frequency $\sim 7\,{\rm Hz}$ and is followed by a series of harmonics (see vertical dashed lines in Fig.\,\ref{pic:sc_env_dyn_6}bc). 
The frequency is almost independent on the mass accretion rate.
The frequency of this QPO corresponds to the time interval that is slightly shorter than the free-fall time scale (\ref{eq:P_QPO}).
This discrepancy may be attributed to the non-zero initial velocity of the flow and the non-zero geometric thickness of the accretion disc.
At mass accretion rates exceeding $\sim 5\times 10^{18}\,{\rm g\,s^{-1}}$), we see a transition where the QPOs with a frequency of $\sim 7\,{\rm Hz}$ and its harmonics are gradually replaced by QPOs with a frequency of  $\sim 5\,{\rm Hz}$ and its associated harmonics  (see grey vertical shaded lines in Fig.\,\ref{pic:sc_env_dyn_6}a).
This phenomenon is likely attributable to the diminishing radiative force, which decreases but does not entirely vanish, thereby continuing to influence the flow dynamics even during the intervals between quasi-periodic flares.
The quality factors of the most distinct QPOs at high mass accretion rates reach a value of $\gtrsim 40$.
At relatively low mass accretion rate ($\sim 10^{18}\,{\rm g\,s^{-1}}$, see Fig.\,\ref{pic:sc_env_dyn_6}d), PDS of the mass accretion rate shows a broad hump at high frequencies ($\gtrsim 20\,{\rm Hz}$).
}

The strength of QPOs of the mass accretion rate due to the influence of the radiative force is strongly affected by the inclination of NS magnetic dipole in respect to the accretion disc plane and beam pattern of X-ray radiation produces at the NS surface.
Both inclination of magnetic dipole (see Fig.\,\ref{pic:sc_env_dyn_2}) and beaming of radiative pattern (see Fig.\,\ref{pic:sc_env_dyn_3}) stabilise the mass accretion rate at the stellar surface.
\red{In the case of accretion onto the inclined magnetic dipole, stabilisation of accretion results in desapearance of QPOs from the PDS, but PDS shows broadband high-frequency component (see Fig.\,\ref{pic:sc_PDS_02}).}

Stabilisation of accretion flow for the inclined magnetic dipole is attributed to the partial coverage of NS magnetosphere by the flow (see Fig.\,\ref{pic:sc_env_acc_05}).
Under this condition, a fraction of X-ray photons can freely leave a system without interaction with the magnetospheric flow. 
In addition, the flow covering NS magnetosphere has larger surface density $\Sigma$ in comparison to the case of magnetic dipole aligned with accretion disc.
The higher surface density results in a reduction of the average acceleration due to the radiative force, as described by equation (see equation \ref{eq:a_rad_ave}). 

\red{
Analyzing the relative fluctuations in surface density across the magnetospheric surface (i.e., its local average value given by equation \ref{eq:Sigma_ave} and standard deviation given by equation \ref{eq:sigma_Sigma}), it becomes evident that these fluctuations intensify toward the central object, as illustrated in Fig.\,\ref{pic:sc_env_acc_06}. 
In the case of an inclined magnetic dipole, the strength of these fluctuations also exhibits azimuthal dependence. Specifically, surface density fluctuations are more pronounced towards the edges of the accretion curtain enveloping the magnetosphere, as highlighted in the lower panel of Fig.\,\ref{pic:sc_env_acc_06}.
It is essential to emphasize that these fluctuations in surface density hold the potential to significantly influence the process of pulse formation. 
A thorough exploration of this effect will be the focus of an upcoming publication.
}

\begin{figure*}
\centering 
\includegraphics[width=18.cm]{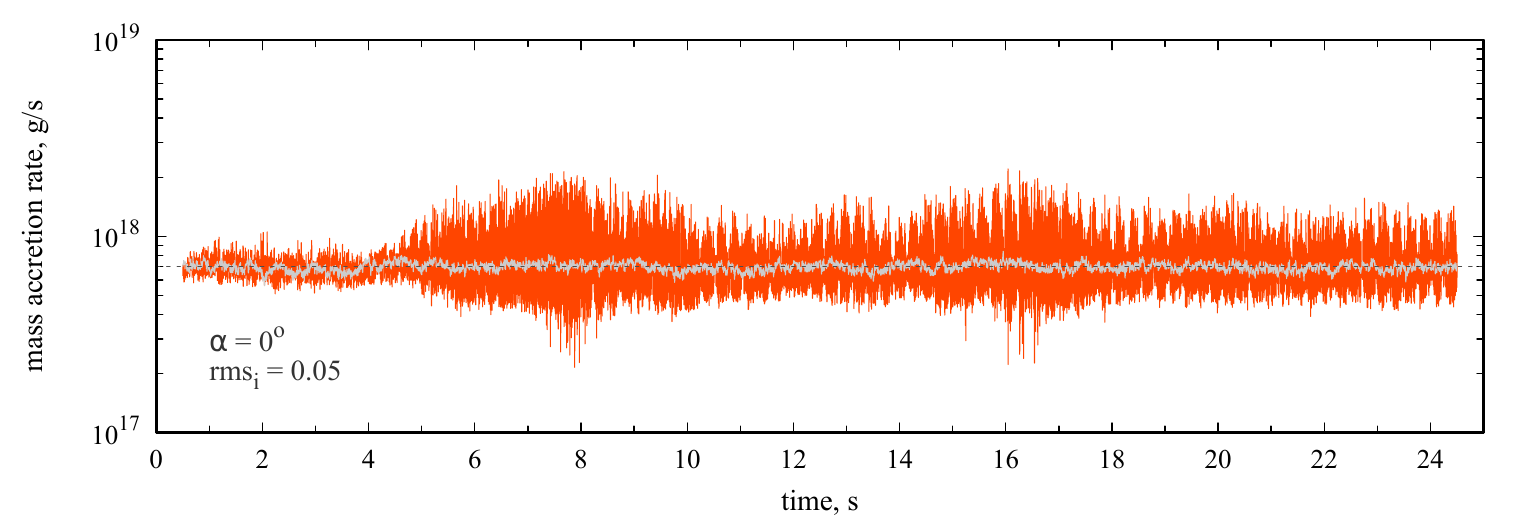} 
\caption{
{
The mass accretion rate onto NS surface \red{(red curve)} calculated for the case of fluctuating mass accretion rate at the inner disc radius \red{(grey curve)}.
The average mass accretion rate at $R_{\rm m}$ is $\dot{M}=7\times 10^{17}\,{\rm g\,s^{-1}}$ \red{(given by horizontal black dashed line)}, ${\rm rms}_i=0.05$.
One can see that fluctuations of the mass accretion rate at the NS surface are sensitive to the mass accretion rate at the inner disc radius. 
It is especially a case when the average mass accretion rate is close to the Eddington limit (as it is here). 
The rotation axis is taken to be orthogonal to the accretion disc plane.
Parameters: $m=1.4$, 
$R_{\rm m}=10^{8}\,{\rm cm}$, $P_{\rm spin}=10\,{\rm s}$, isotropic central source.
}
}
\label{pic:sc_env_dyn_4}
\end{figure*}

\begin{figure*}
\centering 
\includegraphics[width=14.cm]{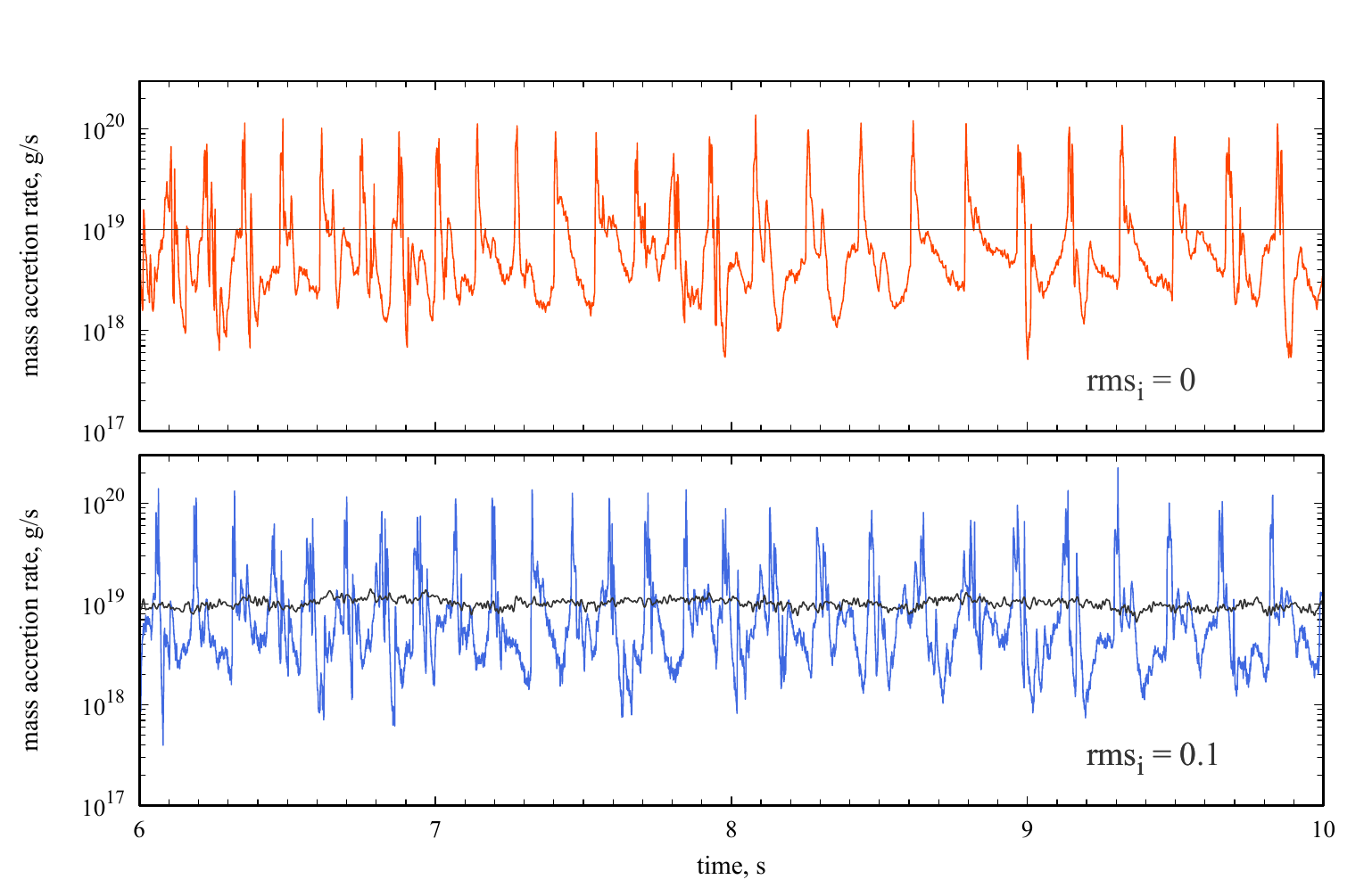} 
\caption{
The red (blue) curve at the upper (lower) panel shows the mass accretion rate at the NS surface calculated for the case of constant (fluctuating) mass accretion rate at the inner disc radius.
The latter is shown with a black line at the upper (lower) panel.
The average mass accretion rate at $R_{\rm m}$ is $10^{19}\,{\rm g\,s^{-1}}$.
In the lower panel, the rms of mass accretion rate fluctuations at $R_{\rm m}$ is taken to be $0.1$.
Parameters: $m=1.4$, 
$R_{\rm m}=10^{8}\,{\rm cm}$, $P_{\rm spin}=10\,{\rm s}$, isotropic central source, $\alpha=0\degr$.
}
\label{pic:sc_env_dyn_5}
\end{figure*}

{
Fluctuations of the mass accretion rate at the inner disc radius (caused by, in particular, the stochastic nature of viscous diffusion) do not affect qualitatively the process of magnetospheric accretion.
The fluctuations at $R_{\rm m}$ increase rms of mass accretion rate variability at the NS surface.
In the case of the average mass accretion rate close to the Eddington value, the fluctuating mass inflow at the magnetospheric boundary might result in variable rms of the mass accreiton rate at the NS surface (see Fig.\,\ref{pic:sc_env_dyn_4}).
It happens because the current mass accretion rate at $R_{\rm m}$ can be occasionally above the Eddington limit due to the fluctuations and then the magnetospheric accretion process turns into the unstable mode. 
At the mass accretion rates much larger than the Eddington limit (we have tested the average mass accretion rate $10^{19}\,{\rm g\,s^{-1}}$), the QPOs are preserved when fluctuations of the accretion rate at the inner disc radius are taken into account (see Fig.\,\ref{pic:sc_env_dyn_5}).
A detailed analysis of accretion timing and calculation of mass accretion rate PDS requires, however, the simulation of long time series and will be discussed in a separate publication. 
}

\red{The initial velocity of the accretion flow at the inner disc radius (within the range $10^6\,{\rm cm\,s^{-1}}\lesssim v_{\rm ini}\lesssim 10^7\,{\rm cm\,s^{-1}}$) has only a negligible impact on the simulated time series of the mass accretion rate and its PDS.
Note that the strength of the correlation between the mass accretion rate onto one and another magnetic pole can be dependent on the details of beam pattern formation and affected by light bending in the vicinity of a NS. 
We are planning to investigate these effects in a separate publication.
}

\begin{figure}
\centering 
\includegraphics[width=8.7cm]{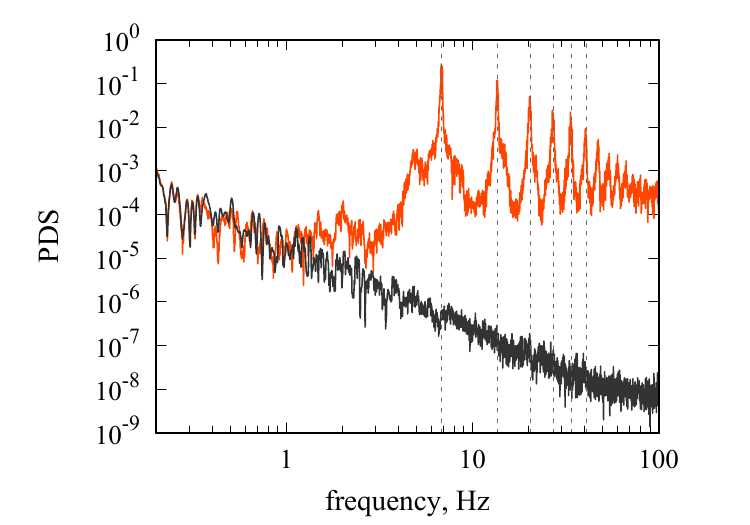} 
\caption{
\red{The modeled PDS of mass accretion rate fluctuations at the inner disc radius (black curve) and at the NS surface (red curve).
One can see QPO at frequency $\sim 6.8\,{\rm Hz}$ and its five harmonics (their freqiencies are shown by vertical dashed lines).
Parameters: $\dot{M}=10^{19}\,{\rm g\,s^{-1}}$, $R_{\rm m}=10^8\,{\rm cm}$.} 
  }
\label{pic:sc_PDS_01}
\end{figure}

\section{Discussion and summary}
\label{sec:Sum}

\red{
We conducted numerical simulations to explore the dynamics of accretion flow over the magnetosphere of bright X-ray pulsars.}
We assumed magnetic field geometry dominated by the dipole component and have taken into account the influence of the gravitational, centrifugal and radiative forces.
The magnetic dipole was taken to be inclined with respect to the plane of the accretion disc, while the rotation axis of a NS was assumed to be orthogonal to the disc, which is a good approximation for the case of accretion onto strongly magnetised NSs, where the surface field strength $B_0\gtrsim 10^{11}\,{\rm G}$. 
\red{These assumptions find support in the determination of geometrical parameters from several XRPs using polarimetric data from the \textit{IXPE} observatory \citep[see e.g.][]{2022NatAs...6.1433D,2023arXiv230206680T,2023arXiv230317325M,2023arXiv230400925M,2023arXiv230515309S}.
The motion of accretion flow along the magnetospheric surface is constrained by the local direction of the magnetic field lines. 
Our calculations were performed under the assumption of opacity being dominated by non-magnetic Compton scattering.}
{Deviations of the magnetic field geometry from the dipole one due to the accretion process \citep{2014EPJWC..6401001L} can affect the dynamics of the magnetospheric accretion flow. 
\footnote{
The degree of magnetic field distortion varies significantly depending on the model assumptions.
In particular, the models proposed by \citealt{1980SvAL....6...14L} and \citealt{1980A&A....86..192A} predict relatively strong distortion, 
while the model by \citealt{1995MNRAS.275..244L} points to small field distortion within the magnetosheric radius.  
}
The calculation of magnetic field geometry perturbed by accretion requires the solution of the RMHD equations, which is beyond the scope of this paper but will be done in the future publications.}

Our examination of the accretion process under low mass accretion rates—where the influence of the radiative force is negligibly small—reaffirms findings previously reported in the literature:
(i) For sufficiently small spin periods and inclined magnetic dipoles, specific regions on the magnetospheric surface exhibit material entrapment due to the interplay between gravitational and centrifugal forces (see Fig.\,\ref{pic:sc_env_acc_05}, see also \citealt{2020MNRAS.496...13A,2023MNRAS.520.4315L}).
(ii) the conditions required for the arising of the ``propeller" effect are dependent on the geometrical thickness of accretion discs and inclination of a magnetic dipole with respect to the disc plane \citep{2022arXiv221112945C}.

\red{
We have demonstrated that accretion flow over the NS magnetosphere is feasible but exhibits instability at super-Eddington mass accretion rates, particularly when the radiative force becomes influential in shaping the flow dynamics (see Fig.\,\ref{pic:sc_env_dyn_1}). 
This instability tends to be quasi-periodic, with a typical period closely aligning with the dynamical timescale at the magnetospheric radius. 
Interestingly, the inclination of the magnetic dipole concerning the disc plane (see Fig.\,\ref{pic:sc_env_dyn_2}) and radiation beaming toward the axis of the magnetic dipole (see Fig.\,\ref{pic:sc_env_dyn_3}) act as stabilizing factors.
Notably, quasi-periodic variability transforms into broadband variability at high Fourier frequencies (see Fig.\,\ref{pic:sc_PDS_02}).
An unstable mass accretion rate onto the NS surface manifests as pronounced fluctuations in the surface density of matter covering the NS magnetosphere (see Fig.,\ref{pic:sc_env_acc_06}). 
Intriguingly, even under conditions of an inclined magnetic dipole and a relatively stable mass accretion rate onto the stellar surface, significant surface density fluctuations persist.
These fluctuations atop a variable mass accretion rate at the NS surface can profoundly impact the pulse profile formation process, rendering pulsations unstable and challenging to detect. 
However, it's essential to note that this assumption requires validation through further numerical simulations.
}

Appearance of high-frequency QPOs \red{and broadband variability} at high mass accretion rates was reported in bright X-ray transient
GRO~J1744-28, where the QPO frequency is $f_{\rm QPO}\simeq 40\,{\rm Hz}$ (see, e.g., \citealt{2019A&A...626A.106M}).
If this QPO is associated with the instability of magnetospheric accretion flow, the inner disc radius is expected to be $\sim 2\times 10^{7}\,{\rm cm}$.
The appearance of high-frequency component ($f>1\,{\rm Hz}$) of PDS at high mass accretion rates was reported in the other Galactic bright X-ray transient - Swift~J0243.6+6124 \citep{2020MNRAS.491.1857D}.

Possibly strong fluctuations of the mass accretion rate can affect the geometry of the emitting region in close proximity to a NS surface: the height of accretion columns \citep{1976MNRAS.175..395B,1988SvAL...14..390L,2015MNRAS.447.1847M} is expected to be dependent on the mass accretion rate. Moreover, according to some theoretical models, variation of the mass accretion rate lead to fluctuating column height on the way to its stable state \citep{2022arXiv220712312A}.
Fluctuating geometry of the emitting region influences the beam pattern \citep{1973A&A....25..233G,2001ApJ...563..289K,2018MNRAS.474.5425M} and formation of pulsation even without accounting for the influence of the magnetospheric accretion flow.

Detection of high-frequency QPOs can be difficult in the case of systems with a strong radiation driven outflows from the disc.
The outflows should collimate X-ray photons from the central object. 
Due to the collimation process, the photons might experience a number of reflections/reprocessings by the walls of accretion cavity.
As a result, one would expect suppression of any variability on time scales smaller than the typical time scale of photon travel inside the cavity \citep{2021MNRAS.501.2424M}.





\section*{Acknowledgements}

AAM thanks UKRI Stephen Hawking fellowship. AI acknowledges support from the Royal Society. VFS thanks Deutsche
Forschungsgemeinschaft (DFG) grant WE 1312/59-1. The work was also 
supported by the German Academic Exchange Service (DAAD) travel grant
57525212 (VFS), the Academy of Finland grants 349144 and 349373, and the V\"ais\"al\"a Foundation (SST).
We our anonymous referee for useful comments and suggestions which helped us improve the paper.

\section*{Data availability}

The calculations presented in this paper were performed using a private code developed and owned by the corresponding author, please contact him for any requests/questions about it. All the data appearing in the figures are available upon request.


\bsp	
\label{lastpage}
\end{document}